\documentstyle{elsart}
\begin{document}
\begin{frontmatter}
\title{A new approach for perovskites in large dimensions}
\author[berlin]{J. Schmalian,}
\author[grenoble]{P. Lombardo,}
\author[grenoble]{ M. Avignon,} and
\author[berlin]{K.- H. Bennemann}
\address[berlin]{Institut f\"ur Theoretische Physik,
 Freie Universit\"at Berlin, Arnimallee 14,
14195 Berlin, Germany}
\address[grenoble]{LEPES-CNRS, BP 166,
38042 Grenoble Cedex 9, France}
%
%\date{\today}
%
%
%\maketitle
%
\begin{abstract}
Using the Hubbard Hamiltonian for transition
metal-3d and oxygen-2p
states with perovskite
geometry, we propose a new scaling procedure for
a nontrivial extension of
these systems  to large spatial dimensions $D$.
The scaling procedure is based on a selective treatment
of different
hopping processes for large $D$ and can not be
 generated by a unique scaling of the hopping element.
The model is solved  in the limit $D \rightarrow \infty$
by the iterated
perturbation theory and using an extended
non-crossing approximation.
We discuss the evolution of quasi particles at the
Fermi-level upon doping,
leading to interesting insight
into the dynamical character of the charge carriers near
 the metal insulator
instability of transition metal oxide systems, three
dimensional perovskites and other strongly correlated
transition metal oxides.
\end{abstract}
\end{frontmatter}
%\vskip 1cm
%\begin{multicol}[2]
The electronic structure of the  strongly correlated copper oxide
superconductors~\cite{BM86}, other
transition metal and rare earth based perovskites~\cite{TOK},
and of  transition metal
oxides~\cite{HUF}  NiO, CoO or MnO  has been an enduring
problem in the last  years, still containing numerous
unresolved questions.
Among them are the applicability of a Fermi liquid
description, the
evolution and doping dependence of coherent quasi
particles near the
Fermi energy in a doped Mott Hubbard or charge
transfer insulator and the
transfer of spectral weight from high to low energy scales.
In these materials, it is of importance
to take the local spin  and charge fluctuations of the
oxygen states
explicitly into account.
A theoretical approach that maintains the dominating
local correlations
and which allows the  calculation of the excitation
spectrum of strongly
 correlated
systems was recently proposed by
Metzner and Vollhard~\cite{MV89}.
They introduced a dynamical mean field theory,
where the system is mapped on
a local problem coupled to an effective
bath.~\cite{MV89,MH89,GK92,PCJ93}
This dynamical mean field approach is exact in
the limit of large
coordination numbers or equivalently  for large spatial
dimensions ($D \rightarrow \infty$).
In most applications of this approach, the one band
 Hubbard model is considered.
A scaling of the hopping element
like $t=t^*/\sqrt{D}$ with fixed $t^*$ leads for
$D \rightarrow \infty$ to a remarkable simplification of the
many body problem while
retaining the nontrivial local dynamics of the
elementary excitations and other main
features of the model.~\cite{GK92,PCJ93}
Only few attempts have been made to extend the
dynamical mean field
theory to systems with more than one orbital
degree~\cite{VG93,GKK93}.

Considering the  $D$-dimensional extension of
the perovskite lattice,
where the TM sites  sit on a  hypercubic lattice and
the oxygen
sites are located between every two nearest neighbour
TM sites,
the following two band Hubbard Hamiltonian for the
TM\,$3d$ and O\,$2p$  orbitals  results:
\begin {eqnarray}
H
&=&\sum_{{\bf k}\sigma}
 \left(\varepsilon_d d^\dagger_{{\bf k} \sigma}
d_{{\bf k} \sigma}
+ \varepsilon_p p^\dagger_{{\bf k} \sigma}
p_{{\bf k} \sigma} \right)
 + \sqrt{2 D} t \sum_{{\bf k}\sigma}
\left( \gamma_{\bf k} d^\dagger_{{\bf k} \sigma}
p_{{\bf k} \sigma}
 + H. c. \right) \nonumber \\
& & +\,U \sum_i d_{i\uparrow}^\dagger d_{i\uparrow}
                  d_{i\downarrow}^\dagger d_{i\downarrow} \,.
\label {Hubb}
\end {eqnarray}
 Here, $d_{{\bf k} \sigma}^\dagger$
($p_{{\bf k} \sigma }^\dagger$) creates
a hole in a TM  (O-2p ) orbital with momentum
${\bf k}$  and spin
$\sigma$. $\varepsilon_d$ and
$\varepsilon_p$ are  the corresponding on-site energy,
 respectively.
$t$ is the amplitude of the nearest neighbour
TM-O hopping integral and $U$
is the Coulomb repulsion of TM holes.
The coherence factor is given by
 $\gamma_{\bf k}^2=1 - \frac{1}{D} \alpha_{\bf k}$
 with $\alpha_{\bf k}= \sum_{\nu=1}^{D} {\rm cos}k_\nu $.
There exists no scaling like $t=t^\ast/D^\beta $ with exponent $\beta$
 leading to a nontrivial limit of Eq.~\ref{Hubb} for $D \rightarrow \infty$.
This results from the  TM coordination number being
 $2D$ whereas the oxygen coordination number is $2$.
A selective consideration of  different hopping  paths on the $D$-dimensional
lattice can be obtained by
the following separate scaling of the
hopping element in the  coherence factor:
$t^2 D \gamma_{\bf k}^2 =
%t ^2 D- t ^2 \alpha_{\bf k}
% \rightarrow
2(t ^*) ^2 -  \sqrt{\frac{2}{D}}(t ^*) ^2 \alpha_{\bf k}$.
Here, the TM-O-TM hopping precesses with identical and
with different TM-sites
are scaled differently such that  for two different TM-sites,
the transfer function
$t^2/(\omega-\varepsilon_p)$ and not $t$ itself is scaled
 similar to  the
one band Hubbard model.
Consequently, the  decoupling of different TM-O "dimers"
of  Ref.~\cite{VG93}
does not occur.
Based on this new scaling procedure, it follows
similar to the one band case, that the self energy
$\Sigma_d(\omega)$ of
TM states
is momentum independent and that the system can
be mapped onto an
Anderson model
with effective hybridization, which has to be calculated
 self consistently.
In the following, we solve the model using the iterated
perturbation
theory~\cite{GK92}
(IPT) and a modified version of the non-crossing
approximation~\cite{KK71,PG89} (NCA).

In Fig. \ref{fig1} we show our results for the TM- and
O-densities of states
obtained within the IPT for half filling $x=0$
($x=n_d+n_p-1$).
 Two Hubbards bands  dominated by TM-states are
 separated by $U$ and two
oxygen dominated bands in the neighborhood of
$\varepsilon_p$ are visible.
Similar to the one band case, the IPT leads to an
 interesting structure  of
the density of states, where now four bands are
ocuring.
However,  the expected insulating behaviour at
half filling (for large $U$)
does not occur.
This results from the overestimation of the spectral weight
of the lower Hubbard band within the IPT by $\propto 1-n_d$.
The occupation of
oxygen sites ($n_p >0$) due to  TM-O hybridization
leads to an overcounting
of copper sites which can be occupied without paying
any Coulomb energy.
Therefore, the  success of the IPT in the one band
case does not
 occur within the two band model.  This is due to the
absence of the
particle-hole symmetry  and the change of the
TM-occupation
number $n_d$ as function of $t$  at half filling.
This shortcoming of the IPT results from the
occupation of O sites and
is expected to vanish for small $n_p$ which occurs
in the limit of
a large value for
 $\Delta=\varepsilon_p-\varepsilon_d$.

However, for the physically interesting
situation \mbox{$\Delta \approx U/2$},
the IPT leads to qualitatively wrong results and
one has to develop
theoretical approaches  which take the local TM-O
many body states
explicitly into account.
Therefore, we extended the NCA , which was
shown to be in excellent
agreement with quantum Monte Carlo (QMC)
 simulations in the
one band case, such that a local
TM-O hybrid  with  16 local eigenstates is coupled
 to an effective
medium. Here, the TM-O singlet and triplet states
which are the important
excitations near half filling are explicitly taken into
account.
In Fig.~\ref{fig2}, we show our results for the
 densities of states obtained
from this hybrid-NCA in the whole energy
range and in the neighborhood
of the Fermi level ($\omega=0$).
One clearly recognizes the two TM-dominated
Hubbard bands and the O-dominated
states near $\varepsilon_p$.
Even the double peak structure of the Hubbard
bands is in qualitative
agreement with  QMC calculations for the
 {\em two dimensional}
three band Hubbard model~\cite{DMH92}.
The O-dominated state at the Fermi level results
 from the formation of a local TM-O (Zhang-Rice)
singlet state~\cite{ZR90}.
The strong temperature dependence of this low
energy singlet state is shown
in  the inset of Fig.~\ref{fig2}, where
the formation of coherent quasi particles on a
temperature scale of
$300\, {\rm K}$ occurs.
This results from the spin fluctuations between
 a singly occupied
TM site and the TM-O singlet state.
Interestingly, these coherent quasi particles occur
on this temperature scale
only for doping values larger than $x \approx 0.04$.
Finally, we indeed find an insulating state for half filling.

In conclusion, we proposed a  new scaling procedure
for perovskite systems
in large spatial dimensions.
It is based on a selective treatment of different
hopping processes and can not be generated
by a unique scaling of the hopping element.
Solving this model within the iterated perturbation
theory
and an extended
non-crossing approximation, it is shown that
this procedure  and the NCA
reproduce important physics of the low  dimensional
situation
including the formation of coherent quasi
particles for low temperatures,
while shortcomings of
the IPT are clearly visible.

Part of the work of J. S. has been done at
LEPES-CNRS, Grenoble. Support by the
 Commission of the European Communities
under EEC-Contract CHRX-CT93-0332
 is gratefully acknowledged.

\newpage
\begin{figure}
\caption{TM (solid line) and O (dashed line)
density of states obtained
within the iterated perturbation theory for half filling. }
\label{fig1}
\end{figure}
\begin{figure}
\caption{TM (solid line) and O (dashed line)
density of states obtained
within the hybrid-NCA for $x=0.1$. The
inset shows the temperature
dependence of the O-DOS for $x=0.08$:
$T=10000$ K (dashed line),
$T=1000$ K (dotted line), and $T=300$ K (solid line).}
\label{fig2}
\end{figure}

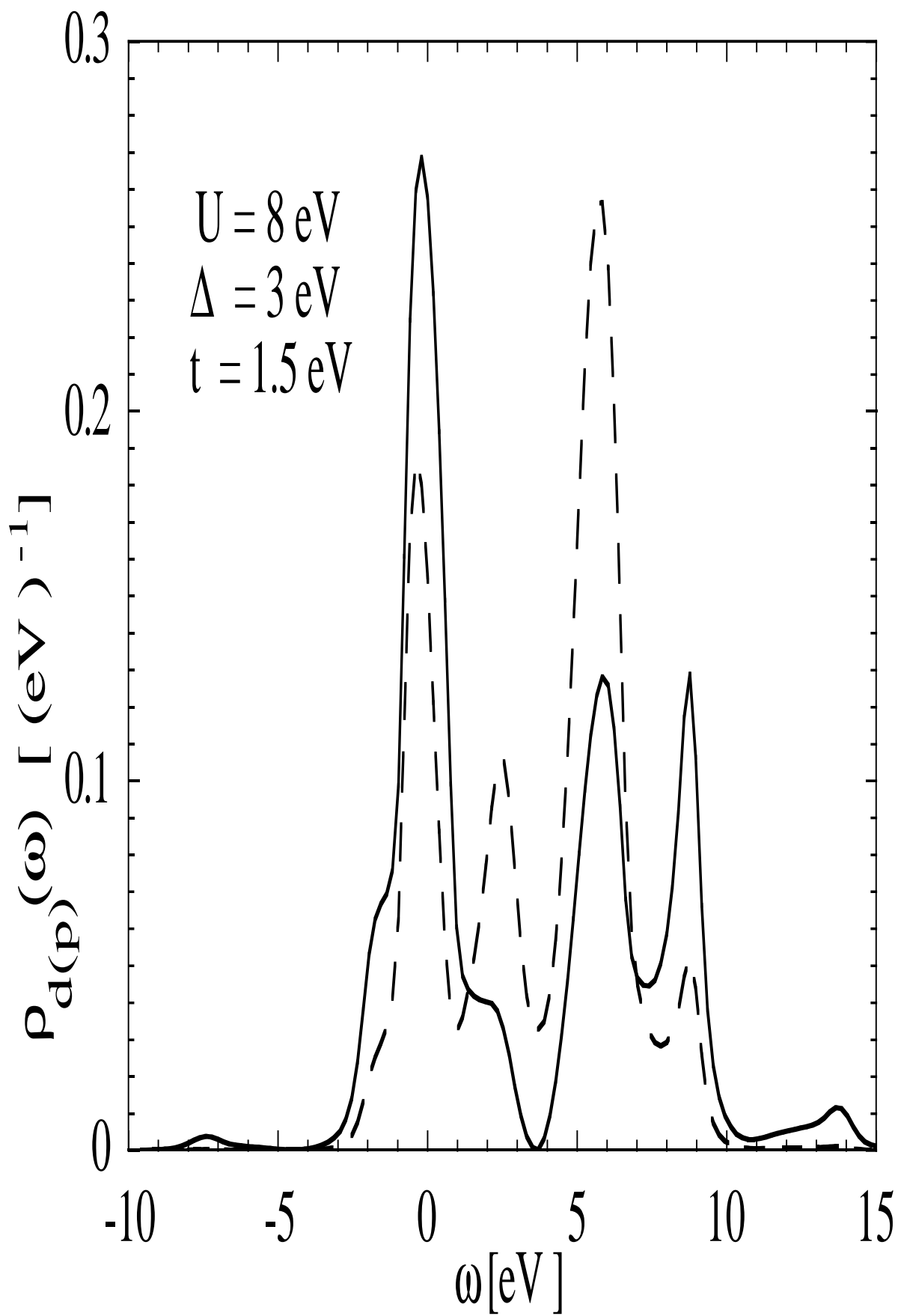
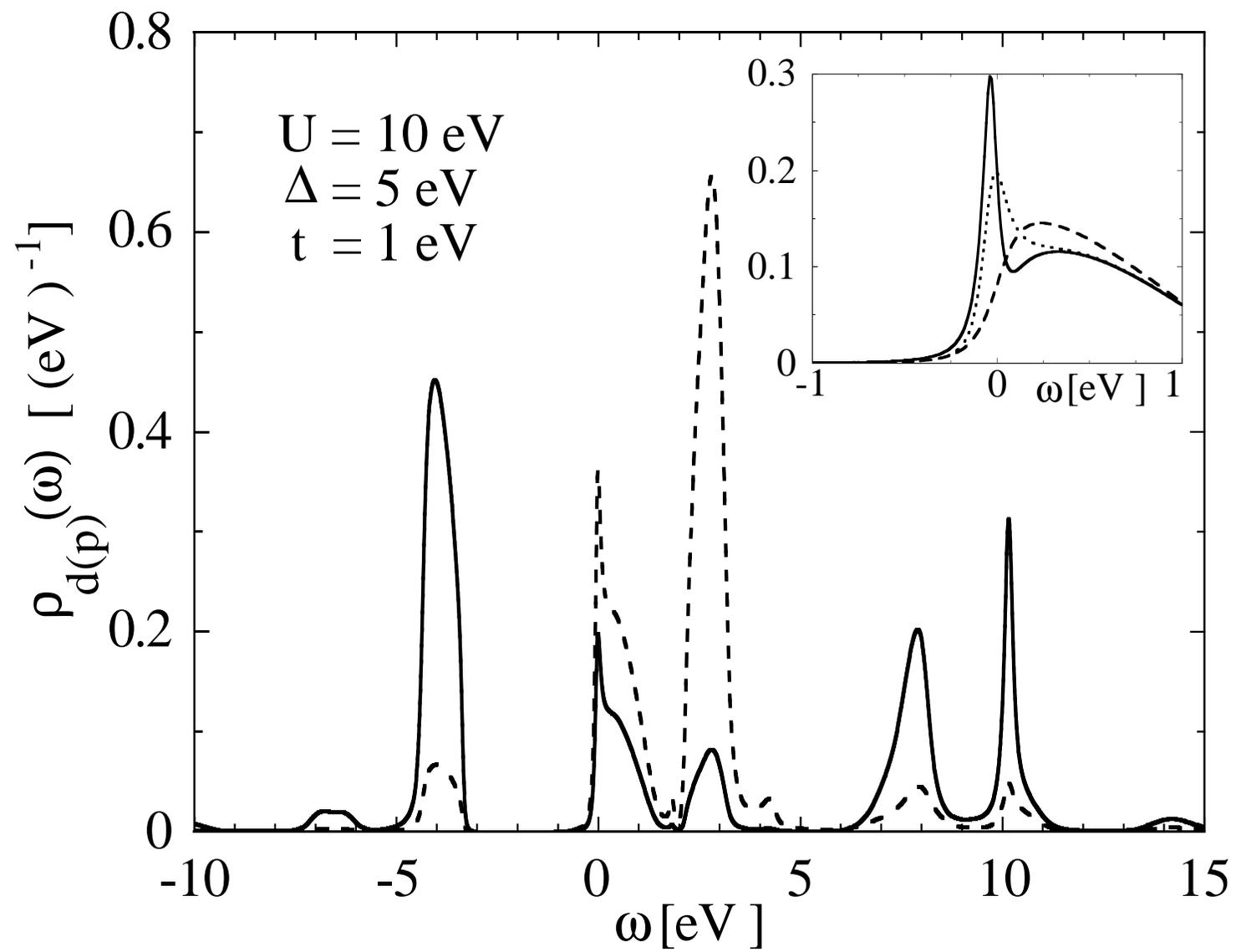
\end{document}